\documentclass[conference]{IEEEtran}
\IEEEoverridecommandlockouts

\usepackage{cite}
\usepackage{amsmath,amssymb,amsfonts,amsthm}
\usepackage{graphicx}
\usepackage{xcolor}
\usepackage{acronym}
\usepackage{algorithm}
\usepackage{algpseudocode}
\usepackage{enumitem}
\acrodef{isac}[ISAC]{integrated sensing and communication}
\acrodef{uav}[UAV]{unmanned aerial vehicle}
\acrodefplural{uav}[UAVs]{unmanned aerial vehicles}
\acrodef{muav}[MUAV]{mission UAV}
\acrodefplural{muav}[MUAVs]{mission UAVs}
\acrodef{cuav}[CUAV]{charging UAV}
\acrodefplural{cuav}[CUAVs]{charging UAVs}
\acrodef{bs}[BS]{base station}
\acrodef{los}[LoS]{line-of-sight}
\acrodef{wpt}[WPT]{wireless power transfer}
\acrodef{soc}[SoC]{state of charge}
\acrodef{rcs}[RCS]{radar cross-section}
\acrodef{pcrb}[PCRB]{posterior Cram\'{e}r--Rao bound}
\acrodef{fim}[FIM]{Fisher information matrix}
\acrodef{mip}[MIP]{mixed-integer program}
\acrodef{madrl}[MADRL]{multi-agent deep reinforcement learning}
\acrodef{u2u}[U2U]{UAV-to-UAV}
\acrodef{ekf}[EKF]{extended Kalman filter}
\acrodef{sca}[SCA]{successive convex approximation}
\acrodef{socp}[SOCP]{second-order cone program}
\acrodef{ao}[AO]{alternating optimization}

\newtheorem{remark}{Remark}

\begin{document}

\title{Toward Mission-Critical ISAC: Reliable Energy-Aware Coordination in UAV Swarms}

\author{
    \IEEEauthorblockN{Masoud Shokrnezhad}
    \IEEEauthorblockA{\textit{ICTFICIAL Oy}\\
        Finland\\
        masoud.shokrnezhad@ictficial.com}
    \and
    \IEEEauthorblockN{Tarik Taleb}
    \IEEEauthorblockA{\textit{Ruhr University Bochum}\\
        Germany\\
        tarik.taleb@ruhr-uni-bochum.de}
    \and
    \IEEEauthorblockN{Amir Javadpour}
    \IEEEauthorblockA{\textit{ICTFICIAL Oy}\\
        Finland\\
        amir.javadpour@ictficial.com}
}

\IEEEaftertitletext{\vspace{-1.5\baselineskip}}
\maketitle

\begin{abstract}
    Unmanned aerial vehicle (UAV) swarms deployed in mission-critical applications must
    simultaneously track a mobile aerial target and maintain reliable data links.
    However, active integrated sensing and communication (ISAC) operation imposes a
    dual energy burden on propulsion and transmission, threatening mission continuity
    through premature battery depletion.
    In this paper, we propose a two-tier UAV swarm architecture in which mission UAVs
    (MUAVs) execute cooperative ISAC for mobile aerial target tracking while dedicated
    charging UAVs (CUAVs), equipped with solar harvesting panels, replenish low-battery
    MUAVs via aerial UAV-to-UAV wireless power transfer (WPT).
    We formulate the joint minimization of the cooperative posterior Cram\'{e}r--Rao
    bound (PCRB) over MUAV trajectories, per-slot sensing-communication time splits,
    WPT scheduling and admission, and CUAV rendezvous trajectories, subject to minimum
    uplink rate, dual-tier energy causality, WPT proximity, collision-avoidance, and
    speed constraints, yielding a non-convex mixed-integer program (MIP) that, to the best of
    our knowledge, is the first to jointly couple cooperative ISAC sensing quality with
    aerial WPT and dual-tier energy management.
    To solve it efficiently, we propose Receding-Horizon Alternating Optimization
    (RHAO), a four-block per-slot algorithm that decomposes the problem into: charging
    admission via the Hungarian algorithm, MUAV trajectory and time-split via successive
    convex approximation (SCA), CUAV rendezvous, and WPT power allocation, with monotone
    convergence guarantees.
    Simulation results demonstrate that RHAO reduces the mean PCRB by $16.8\times$
    over a fixed-time-split baseline and $5.4\times$ over a static-trajectory scheme,
    while the aerial WPT subsystem sustains all MUAVs above the energy-critical
    threshold throughout the full mission horizon.
\end{abstract}

\acused{uav}


\section{Introduction}\label{sec:intro}

Unmanned aerial vehicles (UAVs) are increasingly deployed in mission-critical applications (aerial surveillance,
search-and-rescue, infrastructure inspection, and low-altitude logistics) where
real-time situational awareness and reliable data connectivity must coexist.
\Ac{isac} allows \acp{uav} to share spectrum and hardware resources for simultaneous
cooperative target tracking and telemetry relay, offering high spectral efficiency
without dedicated sensing hardware.
In a multi-\ac{uav} \ac{isac} swarm, each agent contributes range measurements
whose Fisher information combines cooperatively: geometric diversity across the swarm
reduces the \ac{pcrb} on target state estimation far beyond what any single \ac{uav}
can achieve, enabling precise tracking of highly dynamic aerial targets.
However, active \ac{isac} operation imposes a dual energy burden(propulsion for
trajectory following and transmit power for both radar waveforms and uplink data relay) that accelerates battery depletion.
Energy exhaustion forces a \ac{uav} to abort its mission or return to base,
collapsing the cooperative sensing formation precisely when continuous coverage is most
needed.
Two-tier architectures in which dedicated \acp{cuav} carrying solar energy-harvesting
panels autonomously rendezvous with low-battery \acp{muav} and replenish them via aerial
\ac{u2u} \ac{wpt} offer a principled remedy: mission continuity is sustained without
any \ac{muav} leaving the swarm.
Yet, jointly optimizing cooperative \ac{isac} trajectories, per-slot
sensing-communication time splits, \ac{wpt} admission and power scheduling, and
dual-tier energy causality under a unified sensing-quality objective remains an open problem that is the focus of this paper.

Multi-\ac{uav} \ac{isac} for aerial target tracking has attracted growing research
attention.
Yan \textit{et al.}~\cite{yan_curriculum-guided_2026} proposed a system in which a ground \ac{bs}
and multiple \acp{uav} cooperatively minimize the \ac{pcrb} subject to per-\ac{uav}
communication constraints, solving the joint trajectory--beamforming problem via a
curriculum-guided heterogeneous-agent proximal policy optimization algorithm.
Wang \textit{et al.}~\cite{wang_isac-enabled_2026} addressed the same \ac{pcrb} minimization goal
in the low-altitude economy by jointly optimizing \ac{uav} trajectories, bandwidth
allocation, and \ac{uav}--\ac{bs} association through an efficient iterative
descent-direction scheme.
Li \textit{et al.}~\cite{li_cooperative_2026} studied cooperative \ac{isac} for low-altitude
networks with a dynamic time-division strategy, maximizing sum communication rate under
a cumulative radar mutual-information constraint via alternating optimization and \ac{sca}.
Feng \textit{et al.}~\cite{feng_isac-enabled_2025} combined \ac{isac} waveforms with an \ac{ekf}
for 3D urban relay trajectory planning, using a deep Q-network to navigate building
blockages and serve non-\ac{los} users.
None of these works models onboard battery dynamics or aerial \ac{wpt} replenishment.

On the energy management side, Naoya \textit{et al.}~\cite{naoya_multi-uav_2025} studied
multi-\ac{uav} data collection supported by a supply \ac{uav} that charges mission
agents in flight via \ac{wpt}, using simulated annealing and a genetic algorithm to
maximize operational rounds.
Zhang \textit{et al.}~\cite{zhang_eu2u_2025} (\mbox{eU2U}) proposed energy-efficient \ac{u2u}
wireless charging and trajectory co-design for IoT data collection, minimizing total
propulsion and charging-loss energy via a Fermat-point geometric heuristic.
Mondal \textit{et al.}~\cite{mondal_risk-aware_2025} addressed risk-aware \ac{uav}--UGV cooperative
routing under stochastic fuel consumption using an attention-guided reinforcement learning
framework.
Ye \textit{et al.}~\cite{ye_multi-tier_2026}, Gao \textit{et al.}~\cite{gao_-mhappo-enhanced_2025},
and Farhoudi \textit{et al.}~\cite{farhoudi_energy_2026,farhoudi_marl_mec_2026}
examined multi-tier \ac{uav} edge computing systems under energy harvesting, using
Lyapunov optimization and multi-agent reinforcement learning respectively to balance
task delay against long-term energy stability.
None of these works incorporates an \ac{isac} sensing objective or \ac{pcrb} metric.

Despite this progress, a fundamental gap persists across both research threads:
\ac{isac}-\ac{uav} works treat battery energy as unlimited, while
\ac{wpt}-assisted \ac{uav} works focus on data collection or computation without a
cooperative sensing objective.
To the best of our knowledge, no prior formulation simultaneously addresses cooperative \ac{isac} trajectory design,
per-\ac{uav} sensing-communication time splitting, aerial \ac{u2u} \ac{wpt} scheduling
with per-slot admission, and dual-tier energy causality under a unified \ac{pcrb}
criterion.
The \textbf{contributions} of this paper are:
\begin{enumerate}[leftmargin=*,noitemsep,topsep=2pt,label=(\roman*)]
      \item We introduce a \textbf{two-tier swarm architecture} in which \acp{muav}
            execute cooperative \ac{isac} for mobile aerial target tracking while
            solar-equipped \acp{cuav} sustain their operation via on-demand aerial \ac{wpt}.
      \item We formulate the joint optimization of \ac{muav} trajectories,
            sensing-communication time splits, \ac{u2u} \ac{wpt} scheduling and admission,
            and \ac{cuav} rendezvous trajectories under a cooperative \ac{pcrb}
            minimization objective, yielding a non-convex \ac{mip} that jointly couples
            \ac{isac} sensing quality with dual-tier energy causality.
      \item We propose \textbf{RHAO} (Receding-Horizon \ac{ao}), which decomposes the \ac{mip} into per-slot subproblems
            (charging admission, \ac{muav} trajectory and time-split,
            \ac{cuav} rendezvous, and \ac{wpt} power allocation) with a
            monotone convergence guarantee.
\end{enumerate}
The remainder is as follows.
Section~\ref{sec:problem} presents the system and formulates the problem.
Section~\ref{sec:method} describes RHAO.
Section~\ref{sec:simulations} reports simulation results.
Section~\ref{sec:conclusion} concludes.

\section{System Model and Problem Formulation}\label{sec:problem}

\subsection{Network Architecture}\label{sec:network_architecture}

\begin{figure}[t]
    \centering
    \includegraphics[width=0.8\columnwidth]{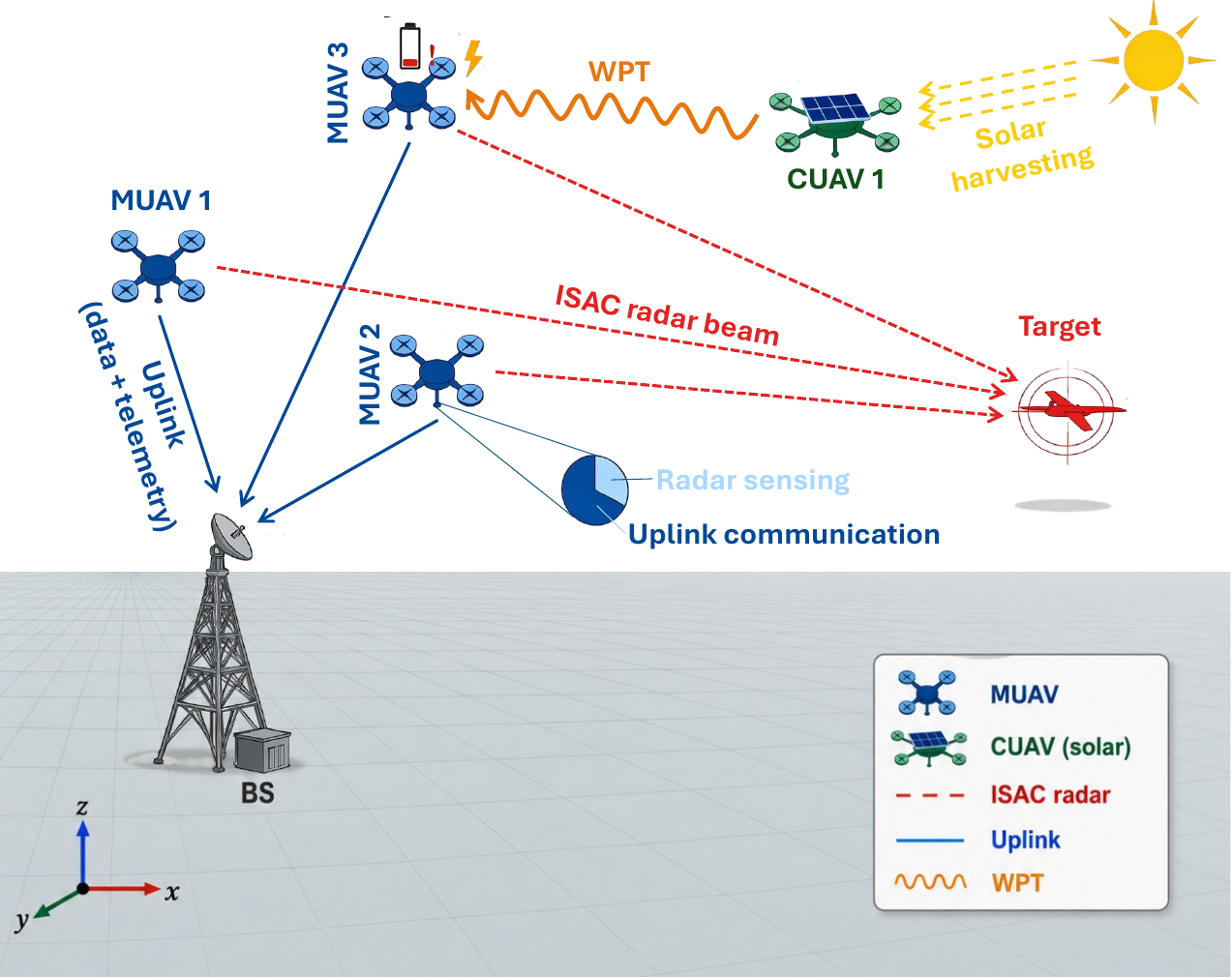}
    \caption{Two-tier \ac{uav} swarm architecture. \acp{muav} (blue) perform cooperative
        \ac{isac}: they transmit radar waveforms toward the mobile target and relay
        sensing data to the \ac{bs} via uplink. \acp{cuav} (green) carry solar harvesting
        panels and recharge low-battery \acp{muav} via aerial \ac{wpt}.}
    \label{fig:system}
    \vspace{-1em}

\end{figure}

We consider a two-tier \ac{uav} swarm comprising $N$~\emph{\acp{muav}},
indexed by $\mathcal{N} = \{1,\ldots,N\}$, and $M$~\emph{\acp{cuav}},
indexed by $\mathcal{M} = \{1,\ldots,M\}$.
The swarm operates cooperatively with a fixed ground \ac{bs} at
$\mathbf{q}_{\mathrm{bs}} \in \mathbb{R}^{3}$ to track a mobile aerial target.
The mission horizon is discretized into $T$ slots of duration $\tau$\,[s],
with $\mathbf{q}_{n}[t] \in \mathbb{R}^{3}$ and $\mathbf{u}_{m}[t] \in \mathbb{R}^{3}$
denoting the 3D positions of \ac{muav}~$n$ and \ac{cuav}~$m$ at slot~$t$, respectively. \acp{muav} execute a cooperative \ac{isac}
mission: they transmit an \ac{isac} waveform to collect range measurements of a
target and relay the resulting sensing/telemetry data to the \ac{bs}.
\acp{cuav} carry solar energy harvesting panels; when a \ac{muav}'s \ac{soc}
falls below a safe level, a \ac{cuav} rendezvous with it and delivers energy via
aerial \ac{wpt}.
Fig.~\ref{fig:system} illustrates the two-tier architecture.

\subsection{\ac{isac} Sensing and Communication Model}

\subsubsection{Target Dynamics}
The target state at slot~$t$ is
$\boldsymbol{\zeta}[t] = \bigl[\boldsymbol{\xi}[t]^{T},\; \dot{\boldsymbol{\xi}}[t]^{T}\bigr]^{T} \in \mathbb{R}^{6}$,
where $\boldsymbol{\xi}[t] = [x_{t},y_{t},z_{t}]^{T}$ is the 3D target position and
$\dot{\boldsymbol{\xi}}[t] = [\dot{x}_{t},\dot{y}_{t},\dot{z}_{t}]^{T}$ is the
corresponding velocity.
Under a near-constant-velocity model over slot duration $\tau$:
\begin{equation}
    \boldsymbol{\zeta}[t]
    = \mathbf{F}\,\boldsymbol{\zeta}[t-1] + \boldsymbol{\varpi}[t-1],
    \quad
    \mathbf{F} = \mathbf{I}_{3} \otimes \begin{bmatrix}1 & \tau \\ 0 & 1\end{bmatrix},
    \label{eq:state_evo}
\end{equation}
where process noise $\boldsymbol{\varpi}[t] \sim \mathcal{N}(\mathbf{0}, \mathbf{Q})$,
with $\mathbf{Q} \in \mathbb{R}^{6 \times 6}$ denoting the process-noise covariance
matrix that captures target maneuver uncertainty between slots.

\subsubsection{Sensing-Communication Time Split}
Each slot, \ac{muav}~$n$ allocates fraction $\delta_{n}[t] \in [0,1]$ to radar
sensing and the complementary fraction $(1-\delta_{n}[t])$ to uplink
communication with the \ac{bs}; both phases use transmit power $P_{n}^{\mathrm{tx}}$.

\subsubsection{Radar Measurement Model}
\ac{muav}~$n$ performs monostatic range sensing of the target.
Let $d_{n}[t] = \|\mathbf{q}_{n}[t] - \boldsymbol{\xi}[t]\|$ denote the true
\ac{muav}--target range at slot~$t$.
The noise variance of the corresponding range measurement, derived from the
radar range equation~\cite{yan_curriculum-guided_2026}, is:
\begin{equation}
    \sigma_{n}^{2}[t]
    = \frac{\kappa_{s}\, d_{n}[t]^{4}}
    {\delta_{n}[t]\,P_{n}^{\mathrm{tx}}},
    \label{eq:range_crb}
\end{equation}
where $\kappa_{s}$ absorbs wavelength, bandwidth, antenna gain, target
\ac{rcs}, noise figure, and noise power~\cite{yan_curriculum-guided_2026}.
Moreover, the range measurement Jacobian with respect to the target state is:
\begin{equation}
    \mathbf{h}_{n}[t]
    = \left[\frac{\bigl(\mathbf{q}_{n}[t] - \boldsymbol{\xi}[t]\bigr)^{T}}
        {d_{n}[t]},\;
        \mathbf{0}_{1\times 3}\right] \in \mathbb{R}^{1\times 6}.
    \label{eq:jacobian}
\end{equation}

\subsubsection{\ac{pcrb}}

The per-slot Fisher information contribution from \ac{muav}~$n$ is:
\begin{equation}
    \mathbf{J}_{M,n}[t]
    = \frac{1}{\sigma_{n}^{2}[t]}\,\mathbf{h}_{n}[t]^{T}\mathbf{h}_{n}[t].
    \label{eq:fim_single}
\end{equation}
Let $\mathbf{J}[t] \in \mathbb{R}^{6 \times 6}$ denote the cooperative \ac{fim}
at slot~$t$, whose inverse $\mathbf{J}[t]^{-1}$ is
the \ac{pcrb} on $\boldsymbol{\zeta}[t]$, as the lower bound on estimation error. $\mathbf{J}[t]$
is updated recursively as \cite{wang_isac-enabled_2026}:
\begin{equation}
    \mathbf{J}[t]
    = \underbrace{\Bigl(\mathbf{F}\,\mathbf{J}[t-1]^{-1}\mathbf{F}^{T} + \mathbf{Q}\Bigr)^{-1}}_{\mathbf{J}_p[t]}
    + \sum_{n=1}^{N} \mathbf{J}_{M,n}[t],
    \label{eq:pcrb_update}
\end{equation}
where $\mathbf{J}_p[t]$ is the \emph{prediction \ac{fim}} propagated from slot~$t{-}1$.
with $\mathbf{J}[0] = \mathbf{J}_{0}$ as the prior \ac{fim}.
The sensing performance metric at slot~$t$ is the trace of the \ac{pcrb} matrix:
\begin{equation}
    \rho_{t} = \mathrm{tr}\!\left(\mathbf{J}[t]^{-1}\right).
    \label{eq:pcrb_metric}
\end{equation}
Note that $\rho_{t}$ couples \ac{muav} positions, sensing time fractions, and
transmit powers through~\eqref{eq:range_crb}--\eqref{eq:fim_single},
and that geometric diversity among \acp{muav} (angular spread around the target)
directly reduces $\rho_{t}$ via the \ac{fim} structure.

\subsubsection{Uplink Communication Model}

The \ac{los}-dominant air-to-ground channel gain from \ac{muav}~$n$ to the \ac{bs} is:
\begin{equation}
    h_{n}[t] = \frac{\beta_{0}}{\|\mathbf{q}_{n}[t] - \mathbf{q}_{\mathrm{bs}}\|^{2}},
    \label{eq:a2g_channel}
\end{equation}
where $\beta_{0}$ is the reference channel gain at 1\,m.
The achievable uplink rate during the communication phase is:
\begin{equation}
    R_{n}[t] = (1-\delta_{n}[t])\, B \log_{2}\!\left(1 + \frac{P_{n}^{\mathrm{tx}}\,h_{n}[t]}{\sigma^{2}}\right),
    \label{eq:rate}
\end{equation}
where $B$ is the allocated bandwidth and $\sigma^{2}$ is noise power.

\subsection{Energy and Wireless Power Transfer Model}

\subsubsection{Propulsion Power}

Both \acp{muav} and \acp{cuav} are rotary-wing \acp{uav}.
The propulsion power as a function of horizontal speed $v$ follows the
standard model \cite{zeng2019uav_energy}:
\begin{align}
    P^{\mathrm{prop}}(v)
     & = P_{0}\!\left(1 + \frac{3v^{2}}{V_{\mathrm{tip}}^{2}}\right)
    + P_{1}\!\sqrt{\sqrt{1+\frac{v^{4}}{4v_{h}^{4}}} - \frac{v^{2}}{2v_{h}^{2}}} \nonumber \\
     & \quad + \tfrac{1}{2}\,d_{0}\,\rho_{a}\,s_{r}\,A\,v^{3},
    \label{eq:prop_power}
\end{align}
where $P_{0}$, $P_{1}$, $V_{\mathrm{tip}}$, $v_{h}$, $d_{0}$, $\rho_{a}$,
$s_{r}$, and $A$ are aerodynamic constants.

\subsubsection{\ac{muav} Battery Dynamics}

Let $B_{n}[t]$ denote the \ac{soc} of \ac{muav}~$n$ at the start of slot~$t$.
It evolves as:
\begin{align}
    B_{n}[t+1]
     & = B_{n}[t]
    - \!\left(P^{\mathrm{prop}}(v_{n}[t]) + P_{n}^{\mathrm{tx}}\right)\!\tau \nonumber \\
     & \quad
    + \zeta\sum_{m \in \mathcal{M}} a_{n,m}[t]\,p_{n,m}[t]\,\tau,
    \label{eq:muav_battery}
\end{align}
where $v_{n}[t] = \|\mathbf{q}_{n}[t+1]-\mathbf{q}_{n}[t]\|/\tau$ is the
\ac{muav} speed, $a_{n,m}[t] \in \{0,1\}$ is the \emph{charging admission indicator}
(\ac{cuav}~$m$ charges \ac{muav}~$n$), $p_{n,m}[t] \geq 0$ is the \ac{wpt} transmit power
from \ac{cuav}~$m$, and $\zeta \in (0,1)$ is the end-to-end \ac{wpt} efficiency
(accounting for beamforming, conversion, and path losses at proximity
range~$d_{\mathrm{wpt}}$).

\subsubsection{\ac{cuav} Battery Dynamics}

Let $B_{m}^{c}[t]$ denote the \ac{soc} of \ac{cuav}~$m$.
\acp{cuav} harvest ambient energy (e.g., solar irradiance) at power $e_{m}[t] \geq 0$
and spend energy on propulsion and \ac{wpt} transmission:
\begin{align}
    B_{m}^{c}[t+1]
     & = B_{m}^{c}[t]
    + e_{m}[t]\,\tau
    - P^{\mathrm{prop}}(v_{m}^{c}[t])\,\tau \nonumber \\
     & \quad
    - \sum_{n \in \mathcal{N}} a_{n,m}[t]\,p_{n,m}[t]\,\tau,
    \label{eq:cuav_battery}
\end{align}
where $v_{m}^{c}[t] = \|\mathbf{u}_{m}[t+1]-\mathbf{u}_{m}[t]\|/\tau$.

\subsection{Problem Formulation}

The optimization problem over the joint variable set
$\boldsymbol{\Phi} = \{\mathbf{q}_{n}[t],\;\mathbf{u}_{m}[t],\;\delta_{n}[t],\;a_{n,m}[t],\;p_{n,m}[t]\}_{n,m,t}$
is:
\begin{subequations}
    \label{eq:P}
    \begin{align}
         & \mathbf{(P):}\;\underset{\boldsymbol{\Phi}}{\min}\quad
        \frac{1}{T}\sum_{t=1}^{T} \rho_{t}\label{eq:obj}                    \\[-2pt]
         & \textit{s.t.} \notag                                             \\[-2pt]
         & R_{n}[t] \geq R_{\mathrm{th}},
        \quad \forall n,\,t,
        \tag{C1}\label{C1}                                                  \\[2pt]
         & 0 \leq B_{n}[t] \leq B_{n}^{\max},
        \quad \forall n,\,t,
        \tag{C2}\label{C2}                                                  \\[2pt]
         & 0 \leq B_{m}^{c}[t] \leq B_{m}^{c,\max},
        \quad \forall m,\,t,
        \tag{C3}\label{C3}                                                  \\[2pt]
         & a_{n,m}[t] \in \{0,1\},
        \quad \forall n,\,m,\,t,
        \tag{C4a}\label{C4a}                                                \\[2pt]
         & \textstyle\sum_{n} a_{n,m}[t] \leq 1,
        \quad \forall m,\,t,
        \tag{C4b}\label{C4b}                                                \\[2pt]
         & \textstyle\sum_{m} a_{n,m}[t] \leq 1,
        \quad \forall n,\,t,
        \tag{C4c}\label{C4c}                                                \\[2pt]
         & 0 \leq p_{n,m}[t] \leq a_{n,m}[t]\,P_{\mathrm{wpt}}^{\max},
        \quad \forall n,\,m,\,t,
        \tag{C5a}\label{C5a}                                                \\[2pt]
         & a_{n,m}[t]\!\left(\|\mathbf{q}_{n}[t]-\mathbf{u}_{m}[t]\|
        - d_{\mathrm{wpt}}\right) \leq 0,
        \;\forall n,\,m,\,t,
        \tag{C5b}\label{C5b}                                                \\[2pt]
         & \|\mathbf{q}_{n}[t+1]-\mathbf{q}_{n}[t]\| \leq v_{\max}^{m}\tau,
        \quad \forall n,\,t,
        \tag{C6a}\label{C6a}                                                \\[2pt]
         & \|\mathbf{u}_{m}[t+1]-\mathbf{u}_{m}[t]\| \leq v_{\max}^{c}\tau,
        \quad \forall m,\,t,
        \tag{C6b}\label{C6b}                                                \\[2pt]
         & \|\mathbf{q}_{n}[t]-\mathbf{q}_{n'}[t]\| \geq d_{\mathrm{safe}},
        \quad \forall n{\neq}n',\,t,
        \tag{C7a}\label{C7a}                                                \\[2pt]
         & \|\mathbf{u}_{m}[t]-\mathbf{u}_{m'}[t]\| \geq d_{\mathrm{safe}},
        \quad \forall m{\neq}m',\,t,
        \tag{C7b}\label{C7b}                                                \\[2pt]
         & \delta_{n}[t] \in [0,1],
        \quad \forall n,\,t.
        \tag{C8}\label{C8}
    \end{align}
\end{subequations}

The constraints encode the following requirements.
\textbf{(C1)} guarantees \textit{a minimum uplink rate} $R_{\mathrm{th}}$ for
sensing-data delivery to the \ac{bs} at every slot.
\textbf{(C2)--(C3)} enforce \emph{energy causality} for both tiers: \acp{muav}
must not over-discharge (preserving operational capability) nor overcharge
(preventing thermal damage), and \acp{cuav} must not transmit more energy than
they store.  The upper bound in \eqref{C2} captures thermal charging-rate
limits, complementing the \ac{wpt} power cap in \eqref{C5a}.
\textbf{(C4a)--(C4c)} define \emph{charging admission}: each \ac{cuav} may charge
at most one \ac{muav} per slot (\ref{C4b}) and each \ac{muav} may receive from at most
one \ac{cuav} (\ref{C4c}), forming a per-slot one-to-one assignment.
\textbf{(C5a)--(C5b)} couple \textit{\ac{wpt} scheduling} to proximity: a non-zero
$p_{n,m}[t]$ is only feasible when $a_{n,m}[t]=1$, and charging is only
admitted when the inter-\ac{uav} distance is within the \ac{wpt} range $d_{\mathrm{wpt}}$.
\textbf{(C6a)--(C6b)} and \textbf{(C7a)--(C7b)} enforce speed limits and
\emph{collision-free flight} for all intra-tier \ac{uav} pairs, respectively.
\textbf{(C8)} bounds the sensing-communication time split.

\begin{remark}[Problem Classification]
    \textbf{(P)} is a non-convex \ac{mip} due to
    (i)~binary admission variables in~\eqref{C4a},
    (ii)~non-convex \ac{pcrb} objective in~\eqref{eq:obj} coupling positions and time splits,
    (iii)~bilinear \ac{wpt} coupling in the battery dynamics~\eqref{eq:muav_battery}--\eqref{eq:cuav_battery},
    (iv)~non-convex collision constraints~\eqref{C7a}--\eqref{C7b}, and
    (v)~temporal coupling across slots via~\eqref{eq:pcrb_update} and battery recursions.
\end{remark}

\begin{remark}[Novelty Over Prior Art]
    Existing \ac{isac}-\ac{uav} works \cite{yan_curriculum-guided_2026,wang_isac-enabled_2026} minimize the \ac{pcrb}
    but treat \ac{uav} energy as unlimited.
    \ac{wpt}-assisted \ac{uav} works \cite{naoya_multi-uav_2025,zhang_eu2u_2025} model battery dynamics but
    lack an \ac{isac} sensing objective.
    \textbf{(P)}~jointly optimizes cooperative \ac{isac} trajectory planning
    (through $\mathbf{q}_{n}$, $\delta_{n}$, and the cooperative \ac{pcrb}
    \eqref{eq:pcrb_update}), \ac{u2u} \ac{wpt} scheduling ($p_{n,m}$) with
    per-slot admission ($a_{n,m}$), and dual-tier energy causality
    \eqref{C2}--\eqref{C3} for both mission and charging agents under a
    unified objective.
\end{remark}

\section{Proposed Method}\label{sec:method}

To solve Problem~\eqref{eq:P}, we propose \emph{receding-horizon \acf{ao}} (RHAO): at each slot~$t$, the
current \ac{fim} $\mathbf{J}[t]$, battery states $(B_n[t], B_m^c[t])$, and
the \acf{ekf} one-step-ahead position estimate
$\hat{\boldsymbol{\xi}}[t] = [\mathbf{F}\hat{\boldsymbol{\zeta}}[t{-}1]]_{1:3}$
serve as known initial conditions, and the slot's decision variables
$\boldsymbol{\Phi}[t]$ are obtained by cycling over four tractable blocks
until convergence, using the prediction \ac{fim} $\mathbf{J}_p[t]$ defined
in~\eqref{eq:pcrb_update}.

\subsection{Block~1 --- Charging Admission}\label{sec:block1}

A \ac{muav}~$n$ is declared \emph{energy-critical} when its \ac{soc}
falls below the trigger threshold $B_n^{\mathrm{trig}} \in (0, B_n^{\max})$,
a design parameter set as a fraction of battery capacity to allow
sufficient rendezvous time before depletion; this yields the candidate set
$\mathcal{N}_c[t] = \{n : B_n[t] < B_n^{\mathrm{trig}}\}$.
The binary one-to-one assignment satisfying~\eqref{C4a}--\eqref{C4c}
is obtained as a minimum-cost bipartite matching:
\begin{equation}
    \{a_{n,m}^*\} = \underset{a_{n,m} \in \{0,1\}}{\arg\min}
    \sum_{\substack{n \in \mathcal{N}_c, m \in \mathcal{M}}}
    a_{n,m}\|\mathbf{q}_n[t{-}1] - \mathbf{u}_m[t{-}1]\|,
    \label{eq:assignment}
\end{equation}
with $a_{n,m}=0$ for $n\notin\mathcal{N}_c[t]$, solved in closed form by
the Hungarian algorithm~\cite{kuhn_hungarian_1955}.
Unassigned \acp{cuav} maintain their last position.

\subsection{Block~2 --- MUAV Trajectory and Time-Split}\label{sec:block2}

\subsubsection{Optimal time-split (closed form)}

For fixed \ac{muav} positions, the objective $\rho_t$ is strictly
decreasing in each $\delta_n[t]$: more sensing time yields tighter
range measurements (lower $\sigma_n^2$) and a larger \ac{fim}.
The rate constraint~\eqref{C1} therefore binds at the optimum, giving
\begin{equation}
    \delta_n^*[t]
    = \left[1 - \frac{R_{\mathrm{th}}}
        {B\log_2\!\!\left(1 +
            \dfrac{P_n^{\mathrm{tx}}\,\beta_0}
            {\sigma^2\|\mathbf{q}_n[t]-\mathbf{q}_{\mathrm{bs}}\|^2}
            \right)}\right]_{0}^{1}\!,
    \label{eq:delta_opt}
\end{equation}
where $[\,\cdot\,]_0^1$ denotes projection onto $[0,1]$.
Each \ac{muav} thus allocates the maximum feasible fraction of the slot
to sensing while guaranteeing the uplink data delivery.

\subsubsection{SCA trajectory update}

With $\{\delta_n^*[t]\}$ fixed, Block~2 reduces to
\begin{equation}
    \min_{\{\mathbf{q}_n[t]\}}\; \rho_t
    \quad \text{s.t.}\quad \eqref{C6a},\,\eqref{C7a}.
    \label{eq:P2}
\end{equation}
The non-convexities are the \ac{pcrb} objective and the
collision constraints~\eqref{C7a}.
We resolve both via \acf{sca}, iterating over position iterates
$\{\mathbf{q}_n^{(l)}\}$, where $l$ is the inner \ac{sca} loop iteration at each slot.

\paragraph{PCRB surrogate.}
The objective $\rho_t=\mathrm{tr}(\mathbf{J}[t]^{-1})$ is non-convex in
$\{\mathbf{q}_n\}$.
To make it tractable we replace it with a \emph{first-order Taylor surrogate}
around the current iterate $\{\mathbf{q}_n^{(l)}\}$, which is linear in
$\{\mathbf{q}_n\}$ and therefore convex.
Let $\mathbf{J}^{(l)} = \mathbf{J}_p[t]+\sum_n\mathbf{J}_{M,n}(\mathbf{q}_n^{(l)},\delta_n^*)$,
$\boldsymbol{\Psi}^{(l)} = (\mathbf{J}^{(l)})^{-1}$, and
$\mathbf{e}_n^{(l)} = \mathbf{q}_n^{(l)}-\hat{\boldsymbol{\xi}}[t]$.
The first-order Taylor expansion of $\rho_t$ around $\{\mathbf{q}_n^{(l)}\}$ is
\begin{equation}
    \rho_t \approx \rho_t^{(l)}
    + \sum_{n=1}^{N}
    \bigl(\nabla_{\mathbf{q}_n}\rho_t\bigr)^{\!T}
    \!\bigl(\mathbf{q}_n - \mathbf{q}_n^{(l)}\bigr),
    \label{eq:taylor}
\end{equation}
where $\rho_t^{(l)}$ and $\mathbf{q}_n^{(l)}$ are constants at iteration $l$.
Substituting $\mathbf{w}_n^{(l)} = -\nabla_{\mathbf{q}_n}\rho_t|_{\mathbf{q}_n^{(l)}}$
and absorbing all constant terms into $c^{(l)}$ reduces~\eqref{eq:taylor} to the \ac{pcrb} surrogate:
\begin{equation}
    \tilde{\rho}^{(l)}\!\bigl(\{\mathbf{q}_n\}\bigr)
    = c^{(l)} - \sum_{n=1}^{N}
    \bigl(\mathbf{w}_n^{(l)}\bigr)^{\!T}\!\mathbf{q}_n,
    \label{eq:pcrb_surr}
\end{equation}
where $\mathbf{w}_n^{(l)}$
is the \emph{sensing gradient}; the direction in which moving \ac{muav}~$n$
most reduces $\rho_t$ locally.
To compute $\mathbf{w}_n^{(l)}$, note that only $\mathbf{J}_{M,n}$ depends on
$\mathbf{q}_n$; the chain rule with the identity
$\nabla_{\mathbf{J}}\mathrm{tr}(\mathbf{J}^{-1})=-(\boldsymbol{\Psi}^{(l)})^{2}$
reduces the problem to differentiating
$\mathbf{J}_{M,n}=(c_n/{d_n^{(l)}}^6)\,\mathbf{e}_n^{(l)}{\mathbf{e}_n^{(l)}}^{\!T}$
with respect to $\mathbf{q}_n$.
Applying the product rule and letting
$\mathbf{A}^{(l)}=\bigl[(\boldsymbol{\Psi}^{(l)})^2\bigr]_{1:3,1:3}$
gives:
\begin{equation}
    \mathbf{w}_n^{(l)}
    = \frac{2c_n}{{d_n^{(l)}}^{6}}
    \!\left[\mathbf{A}^{(l)}\mathbf{e}_n^{(l)}
    - \frac{3\,{\mathbf{e}_n^{(l)}}^{\!T}\mathbf{A}^{(l)}\mathbf{e}_n^{(l)}}
    {{d_n^{(l)}}^{2}}\,\mathbf{e}_n^{(l)}\right],
    \label{eq:sensing_grad}
\end{equation}
with $c_n = \delta_n^* P_n^{\mathrm{tx}}/\kappa_s$ and
$d_n^{(l)} = \|\mathbf{e}_n^{(l)}\|$.
Minimizing~\eqref{eq:pcrb_surr} moves each \ac{muav} along
$\mathbf{w}_n^{(l)}$, balancing proximity to the target
(via $d_n^{-6}$) and cooperative geometric diversity (via $\mathbf{A}^{(l)}$).

\paragraph{Collision constraint linearization.}
The non-convex constraint~\eqref{C7a} is replaced by its first-order
lower bound at the current iterate.
Letting $\mathbf{d}_{nn'}^{(l)} = \mathbf{q}_n^{(l)}-\mathbf{q}_{n'}^{(l)}$:
\begin{equation}
    2\,{\mathbf{d}_{nn'}^{(l)}}^{\!T}(\mathbf{q}_n-\mathbf{q}_{n'})
    \geq d_{\mathrm{safe}}^2 + \|\mathbf{d}_{nn'}^{(l)}\|^2,
    \quad \forall\, n \neq n'.
    \label{eq:coll_lin}
\end{equation}
The per-\ac{sca}-iteration sub-problem--linear
objective~\eqref{eq:pcrb_surr}, second-order cone speed
constraints~\eqref{C6a}, and affine collision
constraints~\eqref{eq:coll_lin}--is a \acf{socp}
solvable in polynomial time by standard convex solvers
(e.g., CVX/MOSEK~\cite{grant_cvx_2014}).

\subsection{Block~3 --- CUAV Rendezvous Trajectory}\label{sec:block3}

Each assigned \ac{cuav}~$m$ ($a_{n,m}^*=1$) flies at maximum speed
toward its paired \ac{muav} subject to~\eqref{C6b}:
\begin{equation}
    \mathbf{u}_m^*[t] = \mathbf{u}_m[t{-}1]
    + \min\!\bigl(v_{\max}^c\tau,\,d_{nm}[t]\bigr)
    \,\hat{\mathbf{d}}_{nm}[t],
    \label{eq:cuav_traj}
\end{equation}
where $d_{nm}[t]=\|\mathbf{q}_n[t]-\mathbf{u}_m[t{-}1]\|$ and
$\hat{\mathbf{d}}_{nm}[t]$ is the unit direction vector.
Unassigned \acp{cuav} hover.
Inter-\ac{cuav} conflicts~\eqref{C7b} are resolved by a priority rule:
the \ac{cuav} serving the more energy-critical \ac{muav} proceeds;
the other holds position for one slot.

\subsection{Block~4 --- WPT Power Allocation}\label{sec:block4}

Let $d_{nm}[t] = \|\mathbf{q}_n[t]{-}\mathbf{u}_m^*[t]\|$ and
$\bar{p}_{nm}[t] = \min(P_{\mathrm{wpt}}^{\max},\, B_m^c[t]/\tau)$
denote the inter-pair distance and the battery-capped power ceiling,
respectively.
WPT is active only when admitted and within proximity range;
the power satisfying~\eqref{C3}--\eqref{C5b} is:
\begin{equation}
    p_{n,m}^*[t] = \begin{cases}
        \bar{p}_{nm}[t] & \begin{array}[t]{@{}l@{}}
                              \text{if } a_{n,m}^*[t]=1 \; \& \; d_{nm}[t]\leq d_{\mathrm{wpt}},
                          \end{array} \\[6pt]
        0               & \text{otherwise.}
    \end{cases}
    \label{eq:wpt_power}
\end{equation}

\subsection{Overall Algorithm and Convergence}

Algorithm~\ref{alg:rhao} summarizes RHAO.
The dominant per-slot cost is the \ac{socp} in Block~2: with $3N$
position variables and $\binom{N}{2}$ linearized collision constraints,
each \ac{sca} iteration has complexity $\mathcal{O}(N^{3.5})$; in
practice, the inner loop converges in 5--10 iterations.

\begin{algorithm}[!tb]
    \small
    \caption{RHAO --- Receding-Horizon Alternating Optimization}
    \begin{algorithmic}[1]
        \Require Initial positions $\mathbf{q}_n[0],\mathbf{u}_m[0]$;
        $\mathbf{J}[0] = \mathbf{J}_0$; batteries $B_n[0],B_m^c[0]$;
        $B_n^{\mathrm{trig}}$; tolerance $\epsilon > 0$.
        \For{$t = 1, 2, \ldots, T$}
        \State \textbf{EKF predict (state):}
        $\hat{\boldsymbol{\xi}}[t] \leftarrow [\mathbf{F}\hat{\boldsymbol{\zeta}}[t{-}1]]_{1:3}$.
        \State \textbf{EKF predict (FIM):}
        $\mathbf{J}_p[t] \leftarrow (\mathbf{F}\mathbf{J}[t{-}1]^{-1}\!\mathbf{F}^T\!+\mathbf{Q})^{-1}$.
        \State \textbf{Block~1:} Solve~\eqref{eq:assignment} via Hungarian $\to \{a_{n,m}^*\}$.
        \State \textbf{Block~2:} Initialize $\mathbf{q}_n^{(0)} \leftarrow \mathbf{q}_n[t{-}1]$, $l \leftarrow 0$.
        \Repeat
        \State Compute $\delta_n^*$ via~\eqref{eq:delta_opt} at $\mathbf{q}_n^{(l)}$.
        \State Compute $\boldsymbol{\Psi}^{(l)}, \mathbf{A}^{(l)}, \{\mathbf{w}_n^{(l)}\}$
        via~\eqref{eq:sensing_grad}.
        \State Solve SOCP $(\text{\eqref{eq:pcrb_surr}},\text{\eqref{C6a}},\text{\eqref{eq:coll_lin}})$
        $\to \{\mathbf{q}_n^{(l+1)}\}$.
        \State $l \leftarrow l + 1$.
        \Until{$\lvert\tilde{\rho}^{(l)}-\tilde{\rho}^{(l-1)}\rvert \leq \epsilon$}
        \State $\mathbf{q}_n[t] \leftarrow \mathbf{q}_n^{(l)}$;\;
        $\delta_n[t] \leftarrow \delta_n^*$.
        \State \textbf{Block~3:} Compute $\mathbf{u}_m^*[t]$ via~\eqref{eq:cuav_traj}.
        \State \textbf{Block~4:} Compute $p_{n,m}^*[t]$ via~\eqref{eq:wpt_power}.
        \State Execute decisions; take radar measurements.
        \State Update \ac{pcrb} via~\eqref{eq:pcrb_update};
        update batteries via~\eqref{eq:muav_battery}--\eqref{eq:cuav_battery}.
        \State \textbf{EKF update:} incorporate range
        measurements~\eqref{eq:range_crb} $\to \hat{\boldsymbol{\zeta}}[t]$.
        \EndFor
    \end{algorithmic}
    \label{alg:rhao}
\end{algorithm}

\begin{remark}[Convergence]
    Each \ac{sca} iterate produces a feasible point with
    $\tilde{\rho}^{(l+1)} \leq \tilde{\rho}^{(l)}$;
    since $\tilde{\rho}$ is bounded below by zero, the inner loop
    converges~\cite{razaviyayn_unified_2013}.
    The outer \ac{ao} loop over Blocks~1--4 inherits monotone
    non-increase of $\rho_t$ per outer iteration.
\end{remark}

\section{Simulation Results}\label{sec:simulations}

\subsection{Setup}

We simulate a $1\,\mathrm{km}{\times}1\,\mathrm{km}$ area over $T{=}60$ slots of
$\tau{=}1\,\mathrm{s}$ each, with $N{=}3$ \acp{muav} and $M{=}2$ \acp{cuav} fixed
at altitude $H{=}100\,\mathrm{m}$.
The ground \ac{bs} is placed at the origin.
A single aerial target at altitude $50\,\mathrm{m}$ moves under a near-constant-velocity
model~\eqref{eq:state_evo} with initial position $(300,300)\,\mathrm{m}$ and velocity
$(7,4)\,\mathrm{m/s}$ (process noise $\sigma_q{=}0.5\,\mathrm{m/s}^2$).
\acp{muav} are initialized in an equilateral triangle of radius $150\,\mathrm{m}$
centred on the target's initial position, providing a well-diversified sensing geometry
from the outset.
Key \ac{isac} parameters are: $\kappa_s{=}5{\times}10^{-9}$, $P_n^{\mathrm{tx}}{=}1\,\mathrm{W}$,
$B{=}10\,\mathrm{MHz}$, $\beta_0{=}10^{-4}$, $\sigma^2{=}4{\times}10^{-13}\,\mathrm{W}$,
and $R_{\mathrm{th}}{=}1\,\mathrm{Mbps}$.
Kinematic limits are $v_{\max}^m{=}25\,\mathrm{m/s}$, $v_{\max}^c{=}20\,\mathrm{m/s}$,
and $d_{\mathrm{safe}}{=}20\,\mathrm{m}$.
Each \ac{muav} starts with $B_n^{\mathrm{init}}{=}12\,\mathrm{kJ}$ out of a
$20\,\mathrm{kJ}$ capacity; \acp{cuav} start with $60\,\mathrm{kJ}$ out of $80\,\mathrm{kJ}$
and harvest $200\,\mathrm{W}$ of solar power continuously.
The \ac{wpt} proximity radius is $d_{\mathrm{wpt}}{=}50\,\mathrm{m}$, the maximum
transmit power is $P_{\mathrm{wpt}}^{\max}{=}500\,\mathrm{W}$ at efficiency $\zeta{=}0.8$,
and the energy-critical trigger threshold is $B_n^{\mathrm{trig}}{=}6\,\mathrm{kJ}$.
All results are averaged over 10 Monte-Carlo runs.

\subsection{Baselines}

We compare RHAO against four reference schemes, ordered from structurally closest to
most naive:
\begin{itemize}[leftmargin=*,noitemsep,topsep=2pt]

      \item \textbf{B1-Static (Wang~et~al.~\cite{wang_isac-enabled_2026}):}
            \acp{muav} are held at their initial positions throughout the mission.
            The optimal time split $\delta_n^*$ \eqref{eq:delta_opt} is applied slot-by-slot,
            and \acp{cuav} use nearest-neighbour \ac{wpt} admission.
            This isolates the gain due to \emph{trajectory optimization} in RHAO.

      \item \textbf{B2-Fixed-$\delta$ Greedy (Li~et~al.~\cite{li_cooperative_2026}):}
            Each \ac{muav} flies at maximum speed toward the current target estimate,
            but uses a \emph{fixed} time split $\delta_n{=}0.5$ (uniform time division).
            This isolates the gain due to \emph{joint time-split optimization} in RHAO.

      \item \textbf{B3-Pure Greedy:}
            Each \ac{muav} flies toward the target estimate with the optimal time split
            $\delta_n^*$ and collision-avoidance deflection, but trajectory decisions are
            made greedily per \ac{muav} without SCA.
            This isolates the value of the \emph{convex trajectory subproblem} in RHAO.

      \item \textbf{B4-Random:}
            \acp{muav} select uniformly random headings and speeds; $\delta_n$ is sampled
            from $\mathcal{U}(0,1)$; \ac{wpt} pairs are random.
            This provides a lower bound on performance.

\end{itemize}

\subsection{Results}

\begin{figure*}[t]
      \centering
      \includegraphics[width=\textwidth]{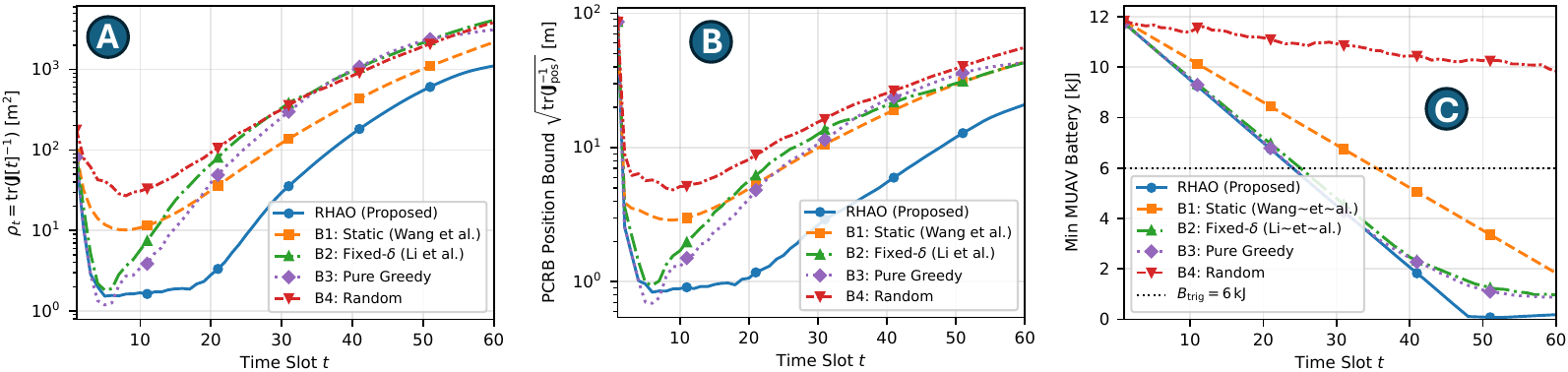}
      \caption{Simulation results averaged over 10 Monte-Carlo runs.
      \textbf{(a)}~Sensing quality: $\rho_t = \mathrm{tr}(\mathbf{J}[t]^{-1})$ vs.\ slot (log scale).
      \textbf{(b)}~Tracking accuracy: PCRB position bound
      $\sqrt{\mathrm{tr}(\mathbf{J}^{-1}_{\mathrm{pos}})}$ vs.\ slot (log scale).
      \textbf{(c)}~Energy sustainability: minimum \ac{muav} battery vs.\ slot;
      dashed line marks the energy-critical trigger $B_n^{\mathrm{trig}}{=}6\,\mathrm{kJ}$
      (Block~1 admission threshold).}
      \vspace{-1em}
      \label{fig:results}
\end{figure*}

\textbf{Sensing quality (Fig.~\ref{fig:results}a).}
All methods begin at the same prior $\rho_1 \approx 80\,\mathrm{m}^2$
(determined by $\mathbf{J}_0$) and improve as measurements accumulate.
RHAO reaches $\rho_t \approx 1\,\mathrm{m}^2$ by slot~5 and sustains sub-$10\,\mathrm{m}^2$
performance throughout most of the horizon, yielding a mean of
$\bar{\rho}^{\mathrm{RHAO}} = 74.5\,\mathrm{m}^2$ (driven by the transient at slot~1
and a late-episode rise as battery constraints bite around slot~50).
\textbf{B1-Static} drops to $\approx 10\,\mathrm{m}^2$ while the target remains near
the initial formation, but degrades monotonically to $\approx 10^3\,\mathrm{m}^2$ as
the target moves away ($\bar{\rho} = 400.6$, $5.4{\times}$ worse than RHAO).
\textbf{B3-Pure~Greedy} achieves a similar low early, but without SCA coordination it
loses angular diversity after roughly 20~slots and degrades sharply
($\bar{\rho} = 644.8$).
\textbf{B2-Fixed-$\delta$} suffers from under-utilized sensing time: fixing
$\delta{=}0.5$ sacrifices up to $40\%$ of the achievable FIM contribution per slot,
resulting in $\bar{\rho} = 1246.8$ ($16.8{\times}$ worse), the worst among all baselines.
\textbf{B4-Random} is close behind ($\bar{\rho} = 1204.6$).

\textbf{Tracking accuracy (Fig.~\ref{fig:results}b).}
The PCRB position bound $\sqrt{\mathrm{tr}(\mathbf{J}^{-1}_{\mathrm{pos}})}$,
where $\mathbf{J}_{\mathrm{pos}}$ is the $3{\times}3$ position-state subblock of
$\mathbf{J}[t]^{-1}$, mirrors the sensing quality trend.
RHAO maintains a mean position RMSE bound of $4.5\,\mathrm{m}$, compared to
$15.4\,\mathrm{m}$ for B1-Static and $22.3\,\mathrm{m}$ for B2-Fixed-$\delta$,
confirming that active tracking and adaptive time-split are both essential.

\textbf{Energy sustainability (Fig.~\ref{fig:results}c).}
RHAO consumes the most energy per slot (active trajectory requires propulsion),
yet \ac{cuav} \ac{wpt} sustains all \acp{muav} above $B_n^{\mathrm{trig}}$
for the $T{=}60$ horizon.
Static and random baselines conserve battery (B1 hovering, B4 low speed)
but sacrifice sensing quality.
This validates receding-horizon optimization with aerial \ac{wpt}:
RHAO achieves the best sensing performance while keeping \acp{muav}
energized for the full mission.

\section{Conclusion}\label{sec:conclusion}

This paper proposed a two-tier \ac{uav} swarm architecture for mission-critical \ac{isac}, where \acp{muav} performed cooperative aerial target tracking while solar-powered \acp{cuav} provided aerial \ac{wpt} to sustain energy continuity.
The joint optimization of trajectories, sensing–communication time splits, and \ac{wpt} scheduling was formulated as a non-convex \ac{mip} minimizing the cooperative \ac{pcrb}, and solved via the proposed RHAO algorithm using \ac{sca} and the Hungarian method.
Simulations showed that RHAO reduced the mean \ac{pcrb} by $16.8\times$ over a fixed-time-split baseline and $5.4\times$ over a static-trajectory scheme, while sustaining all \acp{muav} above the energy-critical threshold throughout the full mission.
Future work will explore extending the framework to multiple simultaneous targets and heterogeneous sensing modalities, as well as incorporating learning-based trajectory policies~\cite{farhoudi_hype_2026} to further reduce per-slot computational overhead.

\section*{Acknowledgment}
This work was, in part, supported by BMFTR, Germany (6GEM+, Grant 16KIS2411), and the EU Horizon Europe programme (6G-Path, Grant 101139172).

\bibliographystyle{IEEEtran}
\bibliography{main.bib}

\end{document}